\documentclass[aps,pra,twocolumn,groupedaddress,showpacs]{revtex4}
\usepackage{graphicx}

\newcommand{\delete}[1]{}

\newcommand{\be}{\begin{equation}}
\newcommand{\ee}{\end{equation}}

\def\beq{\begin{equation}}
\def\eeq{\end{equation}}
\def\bea{\begin{eqnarray}}
\def\eea{\end{eqnarray}}
\def\ba{\begin{array}}
\def\ea{\end{array}}

\begin{document}

\title{Cold atoms as a coolant for levitated optomechanical systems}
\author{Gambhir Ranjit, Cris Montoya, Andrew A. Geraci}
\email[]{ageraci@unr.edu}
\affiliation{Department of Physics, University of Nevada, Reno, NV 89557}

\date{\today}
\begin{abstract}

Optically trapped dielectric objects are well suited for reaching the quantum regime of their center of mass motion in an ultra-high vacuum environment. We show that ground state cooling of an optically trapped nanosphere is achievable when starting at room temperature, by sympathetic cooling of a cold atomic gas optically coupled to the nanoparticle. Unlike cavity cooling in the resolved sideband limit, this system requires only a modest cavity finesse and it allows the cooling to be turned off, permitting subsequent observation of strongly-coupled dynamics between the atoms and sphere. Nanospheres cooled to their quantum ground state could have applications in quantum information science or in precision sensing. 


\end{abstract}


\maketitle

{\section{Introduction}}

While it has been possible to cool individual atoms to their motional quantum ground states for some time \cite{wineland}, a significant amount of research has recently focused on cooling the modes of vibration of larger (mesoscopic) mechanical structures \cite{schwab,heidmann,aspelmeyer,harris,rugarcool,kippenberg,murch,ligogram}. Several such devices have since been cooled to their quantum ground state \cite{cleland,teufel2,painter}. 
These results forge a path towards new hybrid quantum systems where the motion of mechanical resonators can be coupled with other quantum systems, including superconducting qubits \cite{cleland,schwab2,blencowe}, NV centers in diamond \cite{rabl1,kolkowitz}, or the internal or motional states of cold atoms \cite{treutleinbec,pralattice,wang,treutlein2010,
treutlein2011,sympcooltheory}.
In these setups, mechanical oscillators can act as transducers, providing coupling between photons, spins, and charges via phonons \cite{transducers1,transducers2,transducers3,transducers4}. Such transducers could play an important role in quantum networks, by allowing coupling between different types of quantum systems, each with differing advantages.  



Recent experimental results have demonstrated sympathetic cooling of a membrane in a cavity using atoms in a separate vacuum chamber \cite{philippnew}, and theoretical work has shown that by placing a cryogenic membrane in a medium-finesse optical cavity, the strong coupling regime and ground state cooling of the membrane can be achieved via sympathetic cooling of the atoms \cite{sympcooltheory,philippnew}. Sympathetic cooling can significantly enhance optomechanical cooling of mechanical resonators even outside of the resolved sideband regime \cite{mukund}.  Compared with cavity opto-mechanical cooling methods, the sympathetic cooling approach eliminates the requirement of having an ultra-high finesse cavity in a cryogenic vacuum system, and adds the capability of turning off the atomic cooling to observe the subsequent strong-coupling dynamics.

For an atom-membrane coupled system such as that realized in Ref. \cite{philippnew}, ground state cooling can be achieved if the coherent dynamics occur at a sufficiently fast time scale compared with dissipative effects. The fast thermalization rate \beq \Gamma_{\rm{th}}=\frac{k_B T}{\hbar Q}, \label{gth} \eeq where $T$ and $Q$ are the initial temperature and mechanical quality factor of the oscillator, requires cryogenic pre-cooling of the membrane \cite{philippnew}.  On the other hand, because their coupling to the thermal environment is very weak under ultra-high vacuum conditions, the thermalization rate in Eq. (\ref{gth}) for levitated mechanical systems \cite{changsphere,virus,raizen,shortrange,rochester,aspelmeyercavity,ashkin,barkernature,levreview} can be exceedingly low even at room temperature, as the mechanical quality factors are predicted to exceed $Q>10^{11}$. Already impressive quality factors of $10^7$ have been achieved in a moderately high vacuum of $10^{-5}$ mbar \cite{rochester}, and $Q$ is expected to further increase at lower pressures. In this paper we show that optically trapped nanospheres in vacuum can be cooled from room temperature to the ground-state by sympathetic cooling with atoms in an optical lattice. 

For an oscillator with $Q \sim 10^{11}-10^{12}$ at room temperature the thermalization rate in Eq. (\ref{gth}) becomes $\sim 10-100$ Hz.  We calculate that for an ensemble of $5 \times 10^7$ Rb atoms in an optical lattice, the opto-mechanical coupling rate can exceed several kHz, making this system able to attain the strong coupling regime. With a suitable sympathetic cooling rate of order $\sim 10$ kHz, we estimate that ground state cooling of a $300$ nm or smaller diameter sphere is possible. When compared with larger membrane oscillators whose mass can exceed the atomic ensemble mass by a factor of $10^8$ or more, in this setup the collection of Rb atoms has less of an impedance mismatch from the $10^9$ or fewer atoms in the sphere. 

As pointed out in Refs. \cite{cavqed,philippnew,mukund,EPR,zoller2013}, in such a system it may be possible to realize coupling to internal degrees of freedom rather than motional states of the atoms, allowing full quantum state control of the mechanical resonator \cite{law}. Experiments with ground-state cooled levitated mechanical oscillators have an additional versatility, in that by turning the trap off, the quantum wave packet expansion and evolution of the levitated particle in free space can be explored \cite{oriol}. Thus this cooling technique could provide an additional route towards tests of mesoscopic matter-wave phenomena in nanoparticles.  The method also could have more general applicability to the cooling of smaller dielectric particles including clusters of molecules \cite{philippnew,molecules}.


{\section{Sympathetic Cooling of nanoparticles with atoms}}

For concreteness, we imagine the atoms are confined and cooled in a 1-d optical lattice with wavelength $780.74$ nm and linear polarization, red-detuned by $0.5$ nm with respect to the $^{87}$Rb D2 transition. The atom cooling can be performed by molasses cooling \cite{philippnew} or Raman sideband cooling \cite{ramancool}. The atoms can be confined in the transverse directions by an additional far detuned optical lattice, which facilitates the separation of atoms on individual trapping sites. This allows reducing light-assisted collisions during an optical molasses cooling process.  Such light-assisted collisions were shown to limit the cooling performance in recent experimental work on cooling of a membrane using atoms \cite{philippnew}. In Ref. \cite{philippnew}, molasses cooling rates of order $10$ kHz were achieved. In the case of Raman sideband cooling, large numbers of atoms exceeding $10^8$ could be coupled to the sphere \cite{ramancool}.

\begin{figure}[!t]
\begin{center}
\includegraphics[width=1.0\columnwidth]{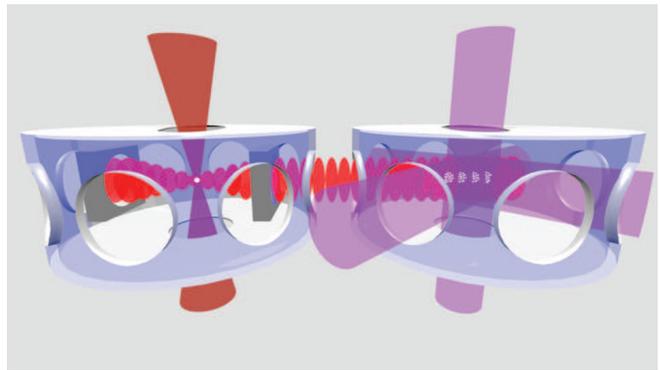}
\caption{(color online) Experimental setup. An ensemble of laser-cooled atoms trapped in an optical lattice (right chamber) is used to sympathetically cool the center of mass motion of a trapped nanobead in a separate optical cavity (left chamber). The left-most mirror of the cavity containing the bead has higher reflectivity to provide increased coupling of the reflected field to the cold atoms. A dual-beam optical tweezer traps the sphere at a position of maximal linear opto-mechanical coupling between the sphere and lattice beam, as discussed in the text. 
\label{sympcoolfig}}
\end{center}
\end{figure}

A separate vacuum chamber containing a $L=5$ cm medium finesse $({\mathcal{F}}=400)$ concentric optical cavity contains a silica nanosphere of radius $a$ trapped by a focused counter-propagating optical tweezer using light of wavelength $\lambda_{\rm{trap}}=2\pi/k_{\rm{trap}}=1550$ nm which is perpendicular to the cavity axis. We assume one mirror of the cavity (farthest from the atoms) has much higher reflectivity, as in Ref. \cite{philippnew}. The $780.74$ nm light propagates between the chambers. The motion of the bead can be measured by imaging the plane of the trap onto a quadrant or split photo-detector. The axial $(z)$ motion of the sphere can also be monitored as phase of the light reflected from the cavity is modulated through the optomechanical coupling.  The proposed experimental arrangement is shown in Fig. \ref{sympcoolfig}.

Using the stronger 1550 nm tweezer, the bead can be positioned at a location of maximal slope of the intensity of the $780.74$ nm cooling laser inside the cavity, to maximize the linear optomechanical coupling. The tweezer intensity can be adjusted to obtain resonance between the sphere motion and atom motion. As the sphere moves, it changes the phase of light reflecting from the cavity. This imparts a force on the atoms, which sit in a potential which depends on the anti-node positions of the standing wave, which are affected by the motion of the sphere. The strength of this coupling is proportional to the ratio of the volume of the bead to the cavity mode volume. Conversely, as the atoms move from their trapping minima, photon momenta are imparted to the atoms to restore them to their equilibrium position, and the resulting intensity of the right or left moving components of the standing wave is increased or decreased, which imparts a force on the trapped bead inside the cavity.  This effect has been observed for atoms previously \cite{treutlein2011,raithel}.

In order for the bead to be initially trapped in high-vacuum conditions, three-dimensional feedback cooling can be employed via position measurements of beads e.g. using split photodiodes \cite{rochester,raizen}. By phase shifting the measured displacement signal and modulating the power of an additional laser, the motion of the bead can be cooled and damped. 

For cavity assisted readout of the displacement of the sphere, the relevant optomechanical coupling is given as $\partial{\omega_c}/\partial{z}= 2 \omega_c g_s/c$. Here $g_s=\frac{3V}{4V_c} \frac{\epsilon-1}{\epsilon+2} \omega_c$, where $\epsilon$ is the dielectric constant of the sphere of volume $V$, and $V_c$ and $\kappa=\pi c/L{\mathcal{F}}$ are the cavity mode volume and linewidth, respectively \cite{changsphere}. Photon shot-noise limits the minimum detectable phase shift to
$\delta \phi \approx 1/(2\sqrt{I})$ where $I \equiv P_c /(\hbar
\omega_c)$ \cite{hadjar}. The corresponding photon shot-noise
limited displacement sensitivity
 is $ \sqrt{S_z(\Omega)}= \frac{\kappa c}{4 \omega_c g_0}
\frac{1}{\sqrt{I}}\sqrt{1+\frac{4\Omega^2}{\kappa^2}}$
\cite{kippenbergdisp}, along the cavity axis for an impedance matched cavity. Here $P_c$ and $\omega_c$ are the laser power and frequency.  For a separate detection laser of $\sim$ 10 $\mu$W with waist 5 $\mu$m, the displacement sensitivity is $2 \times 10^{-14}$ m$/\sqrt{\rm{Hz}}$. This is sufficient to observe the dynamics of the sphere as it is cooled to the ground state since the zero point motion will exceed $2.4 \times 10^{-12}$ m for $a=150$ nm and smaller radius spheres.

Following the discussion of Ref. \cite{sympcooltheory}, the atom-light coupling rate is given by $g_{\rm{at}}=\frac{\omega_{\rm{at}}}{2\alpha k_L \ell_{\rm{at}}} \sqrt{\pi N_{\rm{at}}}$, where $\omega_{\rm{at}}$ is the atom trapping frequency, $\ell_{\rm{at}}$ is the harmonic oscillator length for the atoms, $N_{\rm{at}}$ is the atom number, $k_L=\omega/c$ is the cooling laser wavenumber, and the input laser power is $P=\hbar \omega \frac{\alpha^2}{2\pi}$.  We calculate the sphere-light coupling rate to be
\be g_m = \frac{3}{2} \frac{V}{V_c} \frac{\epsilon-1}{\epsilon+2} \omega k_L \ell_m \frac{\alpha}{\kappa}\frac{1}{\sqrt{\pi}}
  \ee
  where $\ell_m$ is the harmonic oscillator length for the sphere. The effective optomechanical coupling rate connecting the mechanics of the sphere to the mechanical motion of the atoms is given by
   \be
   g=2g_{\rm{at}}g_m=\frac{3}{2}\frac{V}{V_c}(\frac{\epsilon-1}{\epsilon+2})\frac{\omega}{\kappa}\omega_{\rm{at}}\sqrt{\frac{mN_{\rm{at}}\omega_{\rm{at}}}{M \omega_m}},
   \label{geq}\ee where $m$ is the mass of the atom and $M$ is the mass of the sphere. The sphere mechanical resonance is $\omega_m$. For the parameters we consider we are restricted to the weak coupling regime, where $\omega_{\rm{at}},\omega_m>>g^2_{\rm{at}},g_m^2,g_{\rm{at}}g_m$ \cite{sympcooltheory}. In the adiabatic limit where the atom-cooling rate exceeds the atom-membrane coupling rate $g$,
  the sympathetic cooling rate is given by \cite{philippnew} \be \Gamma_{\rm{cool}}=\gamma_{\rm{at}}^{\rm{cool}} \frac{g^2}{\Delta_m^2+(\gamma_{\rm{at}}^{\rm{cool}}/2)^2}
    \ee
    where $\gamma_{\rm{at}}^{\rm{cool}}$ is the atomic cooling rate and $\omega_m$ is bead trap frequency. Here $\Delta_m$ is the difference of the mechanical trapping frequency of the bead and atoms, which we assume to be zero throughout the remainder of the paper. There are also heating terms which lead to momentum diffusion of the atoms and sphere, respectively. They are given by
    \be
    \gamma_{\rm{at}}^{\rm{diff}}=(k_L \ell_{\rm{at}})^2 \gamma_{\rm{se}} \frac{V_0}{\hbar \delta}
     \ee
     for scattering by the atoms, and
     \be
     \gamma_{\rm{sc}}=\frac{2}{5} (\frac{\omega_{\rm{rec,t}}}{\omega_m}) R_{\rm{sc,trap}}+\frac{2}{5} (\frac{\omega_{\rm{rec,L}}}{\omega_m}) R_{\rm{sc,L}}
     \label{scat}\ee  for Rayleigh scattering of trap and lattice laser photons by the sphere \cite{changsphere}. Here $\gamma_{\rm{se}}$ is the spontaneous emission decay rate of the atom, and $V_0$ is the lattice potential depth. $V_0=\frac{\hbar \gamma_{\rm{se}}^2 I_0}{12 \delta I_s}$ for the red-detuned optical lattice, where $\delta$ is the light detuning from the Rb $F=2$ hyperfine component of the ground state \cite{grimm}, $I_0$ is the peak intensity, and $I_s = 1.7$ mW/cm$^2$ is the saturation intensity. The recoil frequencies for the sphere are $\omega_{\rm{rec,t}}=\frac{\hbar k_{\rm{trap}}^2}{2M}$ and $\omega_{\rm{rec,L}}=\frac{\hbar k_{L}^2}{2M}$. The Rayleigh scattering rate $R_{\rm{sc}}=\frac{24\pi^3IV^2}{\lambda^4} \frac{1}{\hbar\omega} (\frac{\epsilon-1}{\epsilon+2})^2$ for light of intensity $I$, wavelength $\lambda$ and frequency $\omega$. $\gamma_{\rm{sc}}$ is dominated by scattering of trap photons for the parameters we consider. Radiation pressure noise also contributes to a momentum diffusion for the sphere given by $\gamma_{\rm{m}}^{\rm{diff}} = 2g_m^2$ \cite{sympcooltheory}.  We assume ideal coupling efficiency to the cavity mode and neglect reduction in light transmission between the vacuum chambers.  The mechanical momentum diffusion rate of the sphere can be significant, and can be reduced by increasing the trap laser wavelength or reducing the size of the nanosphere. The sphere is damped by collisions with the background gas at a rate $\gamma_g = 16P/(\pi \bar{\nu} \rho a)$  where $P$ is the pressure, $\bar{\nu}$ is the gas mean speed, $\rho$ is the density and $a$ is the sphere radius. In sum, the steady state phonon number of the sphere is given by
     \be
     n_{\rm{ss}}=\frac{\gamma_g {\bar{n}}_m+\gamma_m^{\rm{diff}}/2+\gamma_{\rm{sc}}}{\gamma_g+\Gamma_{\rm{cool}}}
     +\left[\frac{\gamma_{\rm{at}}^{\rm{cool}}}{4\omega_{\rm{at}}}\right]^2+\frac{\gamma_{\rm{at}}^{\rm{diff}}}
     {2\gamma_{\rm{at}}^{\rm{cool}}}.     \ee Here ${\bar{n}}_m$ is the mean thermal occupation number of the trapped sphere determined by the initial bath.

\begin{figure}[!t]
\begin{center}
\includegraphics[width=1.0\columnwidth]{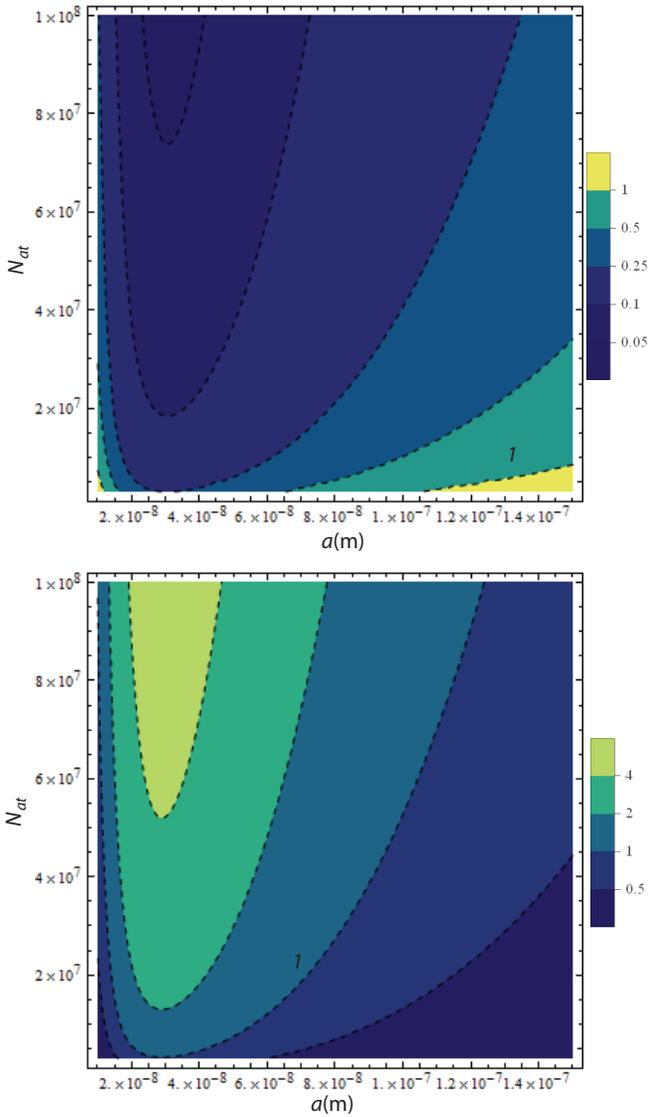}
\caption{Steady-state phonon number in the sphere (upper) and ratio of coupling strength $g$ to the sum of dissipation terms $\gamma_{\rm{at}}^{\rm{diff}}+\gamma_m^{\rm{diff}}+\Gamma_{\rm{th}}+\gamma_{\rm{sc}}$ (lower) versus sphere radius $a$ and number of atoms coupled to the sphere $N_{\rm{at}}$.
\label{nssplot}}
\end{center}
\end{figure}

Experimental parameters appear in Table I for a pressure of $10^{-10}$ torr with $N_{\rm{at}}=5 \times 10^7$. We assume the cavity mode waist is $5$ $\mu$m, the lattice waist is $w_0=30$ $\mu$m, and the lattice beam power is $62$ $\mu$W. The atoms are confined with a lattice depth of $18 E_{\rm{r}}$ where $E_{\rm{r}}=\hbar^2k_L^2/2m$ is the photon recoil energy. The axial and transverse trapping frequencies are $\omega_{\rm{at}}=\sqrt{2V_0k_L^2/m}=45$ kHz and $\omega_r=\sqrt{4V_0/mw_0^2}=263$ Hz, respectively. The optical tweezer has power $460$ mW and the waist near the bead is $2$ $\mu$m.  For an atom cooling rate of $\sim 10$ kHz, quantum ground state cooling of the sphere is achievable for realistic experimental parameters. With the atom cooling turned off, the atom-sphere system evolution can be observed in the strong coupling regime $g> \gamma_{\rm{at}}^{\rm{diff}},\gamma_m^{\rm{diff}},\Gamma_{\rm{th}},\gamma_{\rm{sc}}$ for sufficient $N_{\rm{at}}$ and spheres of sufficiently small size.  A signature of such coupling is the normal mode splitting, which occurs when coherent dynamics occur on a faster time scale than the dissipation in the atoms or mechanics \cite{sympcooltheory,aspelmeyerstrong}. In this case another weakly coupled laser could be used to read out the position spectrum of the sphere in the cavity. Quantum coherent exchange between the atoms and sphere can be studied on the single phonon level. Fig. \ref{nssplot} shows the minimum phonon number as a function of the sphere size and number of atoms coupled to the sphere, along with the ratio $\frac{g}{\gamma_{\rm{at}}^{\rm{diff}}+\gamma_m^{\rm{diff}}+\Gamma_{\rm{th}}+\gamma_{\rm{sc}}}$. Where this ratio is exceeding one, strong coupling dynamics occur.

The dependence of the coupling $g$ in Eq. \ref{geq} is proportional to $\frac{V}{V_c}\frac{\epsilon-1}{\epsilon+2}\sqrt{N_{\rm{at}}m/M}$ where $M$ is the mass of the sphere. For coupling to a membrane oscillator\cite{philippnew}, the dependence on the size of the membrane enters through the term $\sqrt{N_{\rm{at}}m/M}$.  Since the polarizability of the sphere grows as its volume, the larger sphere actually has greater optomechanical coupling even though its mass has a greater mismatch from that of the atomic ensemble. However as indicated in Eq. \ref{scat}, light scattering causes more heating for large spheres, the best strong coupling dynamics are achieved for a certain sphere size, as indicated in Fig. \ref{nssplot} for a fixed cavity mode volume. If the cavity mode volume is scaled with the size of the sphere, then the usual scaling $\sqrt{N_{\rm{at}}m/M}$ valid for membrane oscillators is recovered. However, the cavity mode cannot be arbitrarily made as small as the volume of the nanosphere due to diffraction.

The cavity finesse is chosen to be nearly optimal for our parameters. The finesse of the cavity acts to amplify the effect of the sphere's motion on the phase of the light which returns to the atoms.  However, increasing the finesse much beyond the chosen value will increase the radiation pressure backaction noise $\gamma_{\rm{m}}^{\rm{diff}} \propto {\mathcal{F}}^2$, whereas $g \propto {\mathcal{F}}$. The estimate of $\gamma_{\rm{m}}^{\rm{diff}}$ given in the paper assumes the light in the cavity is on-resonance. In practice, slightly red-detuning the cavity can provide an additional cooling rate to partially counteract this noise, and improve the trapping stability of the bead. However, while operating in the ``bad cavity'' limit ($\kappa>>\omega_m$), such cavity cooling alone is unable to reach the ground state for any value of the detuning \cite{passive}. In the limit of very high finesse, the resolved sideband regime can be obtained, and in this case by red-detuning the light in the cavity, ground state cooling is possible by use of radiation pressure without depending on the atoms.

Laser intensity fluctuations produce variations in the optical trapping frequencies and 
produce a heating mechanism. Fluctuations in the trapping laser frequency can become intensity fluctuations as the light couples into the optical cavity.  
The parametric heating rate due to intensity fluctuations is \cite{lasernoise,lasernoise2} $\Gamma_k=\frac{\omega_m^2}{4}S_k(2\omega_m)$ for a fractional intensity fluctuation spectral density $S_k$. We calculate that a fractional intensity stabilization of $5 \times 10^{-4}$ is required to attain ground state cooling at $45$ kHz. This should be achievable for example using an electro-optic modulator.  Laser pointing fluctuations produce a heating rate $\Gamma_x=\frac{\omega_m^2S_x(2\omega_m)}{4<x^2>}$, where $<x^2>$ is the mean-squared position of the sphere in the trap. The position fluctuation spectral density $S_x$ must be controlled at the level of $10^{-8}$ $\mu$m$^2/{\rm{Hz}}$ to allow ground state cooling. This level of control has been demonstrated previously in other work \cite{lasernoise}.  Finite transmittivity in the optical path $t$ joining the atoms to the sphere and imperfect coupling efficiency to the cavity mode $\eta$ will reduce the sympathetic cooling rate by a factor of $t^2\eta^2$ \cite{philippnew}. $\eta>.75$ and $t>.8$ was achieved in Ref. \cite{philippnew}.

The internal temperature of the sphere $T_{\rm{int}}$ can rise due to absorption of the light from the tweezer and lattice. In ultra-high vacuum, the energy absorbed is re-radiated as blackbody radiation, resulting in a temperature rise of the sphere. We expect $T_{\rm{int}}$ and the center of mass temperature $T_{\rm{CM}}$ are not significantly
coupled over the time scale of the experimental measurements at $P_{\rm{gas}}=10^{-10}$ Torr.
In the case of a sphere with $a<150$ nm, the
blackbody emission is best treated as a point source and the
difference in emitted and absorbed power is given as in Ref \cite{changsphere}.
For fused silica with
dielectric response $\epsilon=\epsilon_1+i\epsilon_2$, with
$\epsilon_1=2$ and $\epsilon_2=1.0 \times 10^{-7}$ as in Ref.
\cite{fusedsilicaloss}, $(\epsilon_{\rm{bb}}-1)/(\epsilon_{\rm{bb}}+2)=0.1$ as in Ref.
\cite{changsphere}, and a bath temperature $T=300$
K, we calculate an internal temperature rise of less than 600 K for the parameters considered.

\begin{table}[!t]
\begin{center}
  \begin{tabular}{@{}cccc@{}}
  \hline
  \hline
  Parameter & 300 nm bead & 100 nm bead  \\
  \hline
$g$ & $2\pi \times (5.9 \times 10^3$ Hz) & $2\pi \times (1.1 \times 10^3$ Hz) \\
$\gamma_{\rm{at}}^{\rm{cool}}$ & $2\pi \times (6.5 \times 10^3$ Hz) & $2\pi \times (1.2 \times 10^3$ Hz) \\
$\gamma_{\rm{m}}^{\rm{diff}}$ &  $2\pi \times (4.5 \times 10^3$ Hz) & $2\pi \times (1.7 \times 10^2$ Hz) \\
$\gamma_{\rm{sc}}$ &  $2\pi \times (6.6 \times 10^3$ Hz) & $2\pi \times (2.4 \times 10^2$ Hz) \\
$\gamma_{\rm{at}}^{\rm{diff}}$ & $2\pi \times (0.27$ Hz) & $2\pi \times (0.27$ Hz)  \\
$\omega_{\rm{at}}, \omega_{\rm{m}} $ & $2\pi \times (4.5 \times 10^4 $ Hz) & $2\pi \times (4.5 \times 10^4 $ Hz) \\
$\Gamma_{\rm{cool}}$ & $2\pi \times (2.1 \times 10^4$ Hz) & $2\pi \times (4.1 \times 10^3$ Hz) \\
$\Gamma_{\rm{th}}$ & $2\pi \times (9$ Hz) & $2\pi \times (28$ Hz) \\
$\gamma_{\rm{g}}$ & $2\pi \times (6.6 \times 10^{-8}$ Hz) & $2\pi \times (1.9 \times 10^{-7}$ Hz) \\
$\kappa$ & $2\pi \times (7.5 \times 10^6$ Hz) & $2\pi \times (7.5 \times 10^6$ Hz) \\
$n_{\rm{ss}}$ & $0.41$ & $0.09$  \\
  \hline
  \hline
  \end{tabular}
\caption{\label{table1} Values for trapping and cooling parameters.}
\end{center}
\end{table}



\section{Comparison with cavity and feedback cooling}

Cavity-cooling in the resolved sideband regime and feedback cooling with sufficiently good measurement imprecision can both also in principle allow ground state cooling of mechanical oscillators \cite{aspelmeyer}. Operation in the resolved sideband regime entails that the mechanical oscillator frequency $\omega_m$ exceeds the cavity linewidth $\kappa$ \cite{passive}. Cavity cooling requires a very high-finesse optical cavity and a sufficiently high mechanical frequency, which could be difficult for levitated microspheres in high vacuum due to heating from optical absorption \cite{barkernature}. Additionally, for experiments with trapped nanospheres, a nebulizer method of trap loading is commonly employed \cite{aspelmeyercavity,rochester}. Particles can become deposited on the cavity mirrors, which can be a challenge for very high finesse cavities. 

A recent paper \cite{bowen} has compared optical feedback cooling of mechanical oscillators to sympathetic cooling using cold atoms. When compared with feedback cooling, it was shown that sympathetic cooling has an advantage for certain parameter ranges. In particular, feedback cooling can reach the ground state even in the bad cavity limit $(\kappa << \omega_m)$ provided the measurement imprecision is sufficiently small. If the position measurement is performed with a measurement cavity of linewidth $\kappa_{MC}$, ground state cooling is possible provided the cooperativity $c_m$ of the mechanical-light interaction exceeds the thermal occupation of the resonator by a factor of 8 \cite{bowen}. The cooperativity $c_m= 4g_0^2 \bar{n}_c/(\Gamma_m \kappa_{MC})$, where $\Gamma_m$ is the mechanical damping rate, $\bar{n}_c$ is the mean intracavity photon number, and the single phonon coupling strength $g_0 = (\partial{\omega_c}/\partial{z}) \ell_{m}$, for ground state oscillator length $\ell_{m}$, describes the shift in the cavity frequency for a given oscillator displacement. In this regime, the mechanical noise spectrum is primarily due to fluctuations in radiation pressure \cite{bowen}.  On the other hand, sympathetic cooling allows ground state cooling even for poor opto-mechanical cooperativity, provided that the atom-light coupling is sufficiently strong. For systems without sufficient cavity quality to perform such position readout of the levitated particle, the sympathetic cooling approach is clearly advantageous.  While there are regions in parameter space where either method dominates, a combined approach yields improved cooling \cite{bowen}. 

 \section{Discussion}

Sympathetic cooling via coupling to cold atoms expands the toolbox for cooling mechanical oscillators to the quantum ground state and can be particularly beneficial in cases where cryogenic operation of the oscillator or operation of very high finesse cavities in the resolved-sideband regime is not practical.
Cooled beads can be used for precision sensing \cite{shortrange,GWprl}, tests of quantum coherence \cite{oriol}, and matter wave interferometry in meso-scale systems \cite{barkerAI,tongcangdiamond,nimmrichter,andyhart}.
 As pointed out in Refs. \cite{EPR,zoller2013,mukund}, by coupling to internal degrees of freedom rather than motional states of the atoms, it is possible to realize a Jaynes-Cummings type system allowing the complete toolbox of cavity QED. As the sphere oscillates inside the cavity, the phase of the light exiting the cavity is modulated due to its motion. A coupling to the atomic internal degrees of freedom can be achieved by transforming this phase modulation into a polarization modulation, which can couple differently to the atoms depending on the atomic internal states. In Ref. \cite{EPR} it was shown that this allows the coupled atom-resonator system to realize an entangled Einstein-Podolsky-Rosen (EPR) state. Furthermore it was pointed out that by coupling to a Rydberg atom, the blockade effect can be used to enable state-transfer, sympathetic cooling, and generation of non-classical states of the mechanical oscillator along with control at the level of single phonons \cite{zoller2013}.

\section{Acknowledgements}
The authors thank Philipp Treutlein and Jonathan Weinstein for useful discussions.  AG is supported in part by grant NSF-PHY 1205994.

\end{document}